# RIS Assisted Wireless Networks: Collaborative Regulation, Deployment Mode and Field Testing

**Yajun Zhao**[1,2]
[1]Beijing Institute of Technology, Beijing 100081, China;
[2]ZTE Corporation, Beijing 100029, China.
Corresponding author: Yajun Zhao (e-mail: zhao.yajun1@zte.com.cn).

*Abstract*—In recent years, RIS has made significant progress in engineering application research and industrialization, as well as in academic research. However, the engineering application research field of RIS still faces several challenges. This article analyzes and discusses the two deployment modes of RIS-Assisted wireless networks, namely Network Controlled Mode and Standalone mode. It also presents three typical collaboration scenarios of RIS networks, including multi-RIS collaboration, multi-user access, and multi-cell coordination, which reflect the differences between the two deployment modes of RIS. The article proposes collaborative regulation mechanisms for RIS and analyzes their applications in the two network deployment modes in-depth. Furthermore, the article establishes simulation models of three scenarios and provides rich numerical simulation results. An actual field test environment is also built, where a specially designed and processed RIS prototype was used for preliminary field test and verification. Finally, this article puts forward future trends and challenges.

## I. INTRODUCTION

To fulfill the requirements of 6G in the future, Reconfigurable Intelligent Surfaces (RIS) stands out as a pivotal potential key technology, offering new physical dimensions, intelligent programmability, and the capability to manipulate electromagnetic waves abnormally [1][2]. RIS has garnered substantial attention from both academia and industry owing to its cost-effectiveness, low power consumption, and straightforward deployment. In recent years, the academic research realm has witnessed breakthroughs and innovations, while the field of engineering application has experienced accelerated progress [3].

Before June 2020, RIS research primarily concentrated in academic circles, with minimal attention directed toward industrial application research. Post-June 2020, a surge in promotional activities related to RIS unfolded in the industry, leading to the establishment of several industrialization organizations. This development significantly propelled the technical research and industrialization of RIS. A pivotal moment occurred in June 2020 when ZTE Corporation, in collaboration with Southeast University and other entities, advocated for the establishment of the "RIS Task Group" within the China IMT-2030 (6G) Promotion Group. This initiative attracted over 30 enterprises and universities, fostering joint efforts to advance the research, standardization, and industrialization of RIS. With the inception of the task group, RIS transitioned from academic prominence to industry-wide recognition [4]. ZTE Corporation initiated the "RIS Research Project" in CCSA TC5-WG6, collaborating with over ten domestic and international enterprises and universities. The IMT-2030 (6G) Promotion Group released the first and second versions of the RIS Technology Research Report during the 6G seminar. ZTE Corporation, Southeast University, and China Unicom jointly hosted the inaugural "THE 1st RIS TECH Forum."[1] In June 2021, ETSI's Industry Specification Group (ISG) on RIS (ISG RIS) gained approval and was subsequently launched in September. The RIS Technology Alliance (RISTA) was established, and the first general meeting of RISTA members took place in Beijing[2]. Sponsored by RISTA and others, ZTE and others jointly hosted the "THE 2nd RIS TECH Forum"[3].

Presently, research teams exhibit an unparalleled level of enthusiasm, with a continuous influx of research findings on RIS. However, existing research predominantly centers on uncomplicated RIS application scenarios involving a single base station/access point. There is a notable dearth of research investment and public achievements pertaining to RIS applications in intricate network scenarios. Regarding RIS deployment, existing work has mainly focused on exploring aspects such as RIS deployment density, deployment location and the numerical configuration of the RIS elements. The paper [5] shows that the plane deployment of RIS should be on the BS-side or the user-side, and the deployment height of the RIS should be close to the other side of the shorter link. The study [6] determined that deploying the RIS near either the Node B (NB) or the User Equipment (UE) is more advantageous when the distance from the BS to the UE exceeds twice the distance from the RIS to the BS-UE line.

Yajun Zhao is with Beijing Institute of Technology, Beijing 100081, China, and with Wireless Product R&D Institute of ZTE Corporation, Beijing 100029, China (e-mail: zhao.yajun1@zte.com.cn).  ORCID: Ya-jun ZHAO, http://orcid.org/0000-0001-8823-5282
Jianwu Dou, and Xiangyang Duan, was with Wireless Product R&D Institute of ZTE Corporation, Shenzhen 518057, China (e-mail: {dou.jianwu, duan.xiangyang}@zte.com.cn).

---

[1] http://2021.risforum.com/
[2] http://www.risalliance.com/LiveVideoServer/risWeb/events_202204_en.html.
[3] http://2022.risforum.com/en/

Conversely, when the distance from the UE to the BS is equal to or less than twice the distance from the RIS to the BS-UE line, optimal deployment of the RIS is in the middle of the BS and the UE. The literature [7]-[9] explores the framework of distributed RIS. In our prior research endeavors, we pioneered the exploration of RIS network coexistence [10][11], with a specific focus on the deployment of RIS in high-speed railway communication scenarios [12].

The choice of deployment mode holds significant importance in RIS research. Building upon our prior investigation [10][11], which examined the network control aspect of RIS, we categorized its deployment mode into two distinct types: Network-Controlled Mode and Standalone Mode. We conducted a preliminary analysis and comparison of these modes. This paper delves deeper into the analysis and discussion of the two deployment modes, encompassing theoretical scrutiny, numerical simulation, and the establishment of an actual field test environment for preliminary comparative field tests. To facilitate clarity in expression, we reiterate the concepts of the two RIS network deployment modes previously outlined. The mode in which RIS is subject to network control is denoted as "*Network-Controlled Mode (NCM)*," while the mode where RIS exercises self-control is referred to as "*Standalone Mode (SAM)*" [13].

This article presents several noteworthy contributions. Firstly, it conducts a comprehensive analysis of RIS network deployment modes. Secondly, it introduces collaborative regulation mechanisms for RIS and thoroughly analyzes their applications in the two types of network deployment modes. Thirdly, the article outlines the construction of an actual field test environment, employing a specially designed and processed RIS prototype for preliminary field testing and verification.

The structure of this article is organized as follows. Section 2 presents the system model of RIS networks, which will be utilized in subsequent simulations and analysis. Section 3 offers an in-depth analysis and comparison of the two deployment modes, NCM and SAM, within three typical scenarios. In Section 4, simulation models for the three scenarios are established, and comprehensive numerical simulation results are provided. Section 5 details the establishment of the actual field test environment and the preliminary testing of the two modes. Section 6 outlines the anticipated trends and challenges for the future. Finally, in Section 7, conclusive remarks are drawn.

## II. SYSTEM MODES AND PROBLEM ANALYSIS

This section will present the network architecture and system model of NCM and SAM, along with an analysis of the issues and challenges encountered by these two models.

### A. SYSTEM MODES

The system architecture of NCM is illustrated in FIGURE 1(a). A control link connection is established between Node B (NB) and RIS. This control link facilitates the exchange of channel state information (CSI) and control signaling between

NB and RIS. The control link connecting NB-RIS may directly interface with RIS, and RISs may or may not be interconnected. The system architecture of SAM is depicted in FIGURE 1(b). In this configuration, the RIS of SAM is deployed independently. There exists no control link between Node B (NB) and RIS, and similarly, there are no control links between individual RISs.

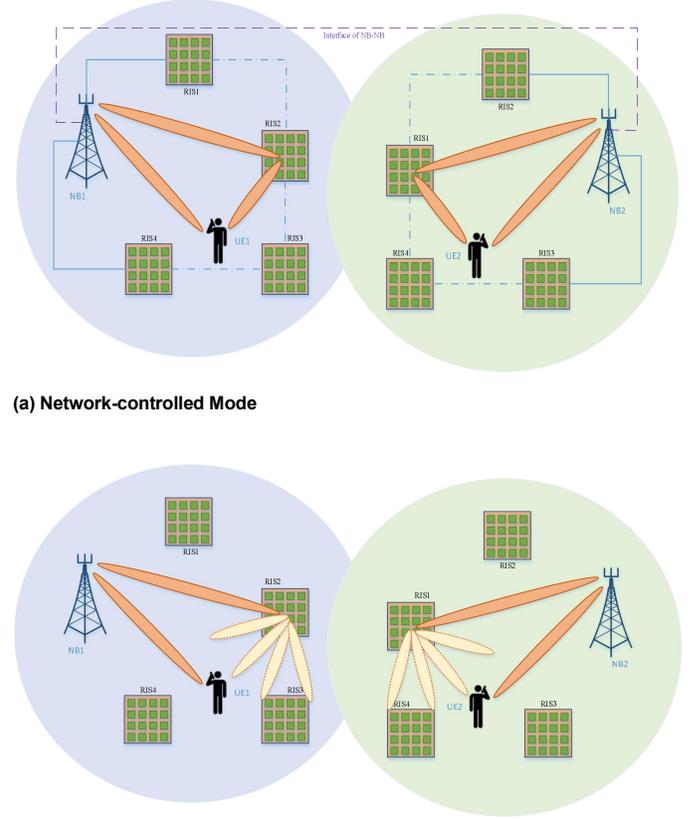

**(a) Network-controlled Mode**

**(b) Standalone Mode**

FIGURE 1. System architecture and model.

The general system model of RIS-assisted wireless network downlink is shown in the formula (1).

$$\mathbf{H}_{DL} = \mathbf{H}_{DL,ris-ue}^{H} \mathbf{\Phi}_{ris,i} \mathbf{G}_{DL,nb-ris} + \mathbf{H}_{DL,direct} \qquad (1)$$

where, $\mathbf{H}_{DL,ris-ue}$ is the downlink channel between RIS and UE; $\mathbf{\Phi}_{ris,i}$ is the regulation matrix used by the $ith$ RIS; $\mathbf{G}_{DL,nb-ris}$ is the downlink channel between NB and RIS; $\mathbf{H}_{DL,direct}$ is the direct downlink channel between NB and UE.

Without losing generality, to simplify the discussion, we assume that the direct path between NB and UE is blocked, and the signal of the direct path can be ignored. Then the system model of the downlink, that is formula (1), is rewritten as formula (2).

$$\mathbf{H}_{DL} = \mathbf{H}_{DL,ris-ue}^{H} \mathbf{\Phi}_{ris,i} \mathbf{G}_{DL,nb-ris} \qquad (2)$$

Based on the characteristics of the system architecture of the above two modes, the differences in system models between the NCM and SAM are mainly reflected in the calculation of the regulation matrix $\Phi_{ris,i}$.

For NCM, due to the existence of interactive links between NB and RIS, when calculating the regulation matrix $\Phi_{ris,i}$, it can not only rely on the capabilities of RIS itself but also be assisted by the network. With the help of NB, RIS can fully obtain the information on the target channel and main interference channels, so that the influence of the target channel and related interference channels can be considered when optimizing the regulation matrix $\Phi_{ris,i}$. That is, NCM regulation/beamforming can achieve cooperative control or centralized control. Then the calculation of the regulation matrix $\Phi_{ris,i,NCM}$ of NCM mode can be expressed as formula (3).

$$\Phi_{ris,i,NCM} = f(\mathbf{H}_{DL,ue_1}, \mathbf{H}_{DL,ue_2}, ..., \mathbf{H}_{DL,ue_i}, ..., \mathbf{H}_{DL,ue_K}) \quad (3)$$

where, $\mathbf{H}_{DL,ue_i} = \mathbf{H}_{DL,ris-ue_i}^H \Phi_{ris,i,ue_i} \mathbf{G}_{DL,nb-ris}$ is the downlink channel between NB and $ue_i$; $K$ is the total number of related UE. The function f(.) herein signifies a general function utilized to compute the RIS tuning matrix. It is employed in this context to denote the essential parameters necessary for calculating the RIS regulation matrix $\Phi_{ris,i,SAM}$.

Regarding SAM, RIS can only rely on its own ability to calculate the regulation matrix. In other words, when calculating its own regulation matrix, the RIS can only rely on the channel information of a single target UE or unknown any channel information. SAM can only adopt independent and distributed control mode, without coordination and interaction with the network or other RISs. It pursues self-optimization and is difficult to achieve joint optimization with its surrounding nodes. Then the calculation of the regulation matrix $\Phi_{ris,i,SAM}$ of NCM mode can be expressed as the formula (4).

$$\Phi_{ris,i,SAM} = f(\mathbf{H}_{DL,ue_i}) \quad (4)$$

### B. PROBLEM ANALYSIS

Many existing RIS studies operate under the assumption of a controller linked to the network, tasked with overseeing the electromagnetic regulation behavior of RIS. This assumption, particularly prevalent in scenarios involving multiple RISs, embodies the ideal NCM model outlined earlier. While NCM yields improved network performance, it introduces complexities in controlling information interaction. To address this complexity, literature [14][15] proposes the adoption of SAM. In [14], it is suggested that RIS can conduct channel estimation and beamforming based on interference phenomena, a mechanism limited to single-user scenarios. Additionally, literature [15] proposes self-regulation and beamforming through a blind beamforming mechanism. However, blind beamforming, reliant on statistical channel information, lacks the effectiveness of beamforming using known Channel State

Information (CSI) and is suitable only for single-user cases. Both interference-based channel estimation and beamforming and blind beamforming mechanisms necessitate RIS awareness of the served UE and the target for optimization, limiting their applicability to single UE scenarios. While these mechanisms endow RIS with distributed computing capabilities, mitigating the need for centralized computing by NB and reducing NB-RIS interaction, they do not eliminate the requirement for RIS-network interaction. Moreover, achieving these goals entails increased RIS complexity, contradicting the fundamental principles of simplicity and cost-effectiveness in RIS design.

Adhering to the fundamental principle of designing RIS with simplicity and cost-effectiveness, we presume that RIS possesses straightforward functionalities. In the case of NCM, aided by NB, RIS acquires ample information concerning target channels and primary interference channels. This enables the consideration of the impact of target channels and related interference channels when optimizing the regulation matrix. The regulation of RIS in NCM can achieve either collaborative control or centralized control. In contrast, SAM can only obtain CSI and compute the regulation matrix independently. Essentially, when calculating its own regulation matrix, RIS relies solely on the CSI of a single target UE or, in some instances, unknown channel information. SAM exclusively adopts independent and distributed control modes, lacking coordination and interaction with the network or other RISs. SAM pursues self-optimization, making it challenging to achieve joint optimization with surrounding nodes. In [14], it is suggested that RIS can achieve channel estimation and beamforming based on interference phenomena; however, this mechanism is limited to single-user scenarios. In [15], a blind beamforming mechanism is proposed for RIS to achieve self-regulation and beamforming.

### III. RIS COLLABLRATIVE REGULATION MECHANISM AND NETWORK DEPLOYMENT MODE

As deduced from the aforementioned analysis, the disparities between the two system architectures primarily manifest in scenarios necessitating collaboration. This section will delve into several typical collaboration scenarios encountered by RIS in practical wireless networks. These scenarios include multi-RIS collaboration, multi-user access, and multi-cell coordination. Furthermore, the scenario of multi-network coexistence, though crucial for RIS in practical wireless networks, has been thoroughly discussed in our prior work [10] and will not be reiterated here.

### A. MULTI-RIS COLLABORATION

In this section, multi-RIS collaboration pertains to the installation of multiple RISs within the coverage area of a cellular cell, coupled with the application of conventional Coordinated Multi-point (CoMP) mechanisms leveraging these RISs. CoMP is a wireless communication technique employed to enhance performance, particularly at the cell edge where interference poses a significant challenge. Similar to the traditional CoMP [16], RIS-based CoMP can also be categorized into two modes:

CoMP-JT: Multiple RIS joint regulation, wherein various signal components of a user terminal (UE) are regulated by different RIS and traverse different propagation paths. This mode can be termed as collaborative multi-RIS joint regulation transmission.

CoMP-DPS: Dynamic selection of one RIS to regulate the UE signal, which can be denoted as collaborative multi-RIS dynamic RIS selection regulation transmission.

This article specifically delves into the mode of RIS-based Coordinated Multi-point Joint Transmission (CoMP-JT) in the presence of phase difference challenges.

Since the advent of the 4G LTE era, CoMP technology has been a subject of discussion and integration into 4G LTE standards, yet widespread commercialization has not been realized. Although CoMP technology was deliberated upon during the development of 5G standards, the current implementation only involves non-coherent CoMP-JT. The predominant benefits of CoMP are derived from coherent CoMP-JT. The gains achieved solely through non-coherent CoMP-JT are limited, posing challenges for broad commercial adoption. The primary hurdle in implementing coherent CoMP with traditional active phased arrays lies in the intricate problem of mutual calibration of antennas. RISs, utilizing passive control and lacking the inherent challenge of channel reciprocity calibration for antennas, offer a viable solution for achieving coherent CoMP-JT. Hence, a practical approach to realize coherent CoMP involves substituting RISs for traditional active phased arrays.

However, achieving RIS-based coherent CoMP-JT faces additional challenges, particularly in addressing synchronization issues during collaborative regulation among multiple RISs for joint transmission. Synchronization deviation arises from the differences in synchronization among crystal oscillators in the control modules of various RISs and the initial phase deviation of the regulation matrix. Notably, the crystal oscillation synchronization deviation of the RIS control module can be readily resolved using traditional methods such as GPS, and hence, this aspect will not be reiterated in this paper. This article focuses on studying the synchronization of RIS regulation stemming from the initial phase deviation of the regulation matrix. Specifically, the initial phase deviation of the regulation matrix may introduce varying phase differences in the regulation among multiple collaborating RISs, potentially affecting the performance of collaborative transmission. Consequently, it becomes imperative to employ effective methods to coordinate phase adjustments among multiple RISs to realize coherent CoMP-JT. Given the current insufficient research focus on this issue, this article delves into possible solutions and mitigation strategies to address this synchronization challenge.

Without loss of generality, taking Downlink (DL) as an example and referring to RIS channel model formula (2), the RIS-based CoMP-JT channel model $\mathbf{H}_{DL,CoMP-JT}$ can be expressed as equation (5).

$$\mathbf{H}_{DL,CoMP-JT} = \sum_{i=1}^{L} \mathbf{H}_i^H \mathbf{\Phi}_{ris,i} \mathbf{G}_i \qquad (5)$$

where, $\mathbf{H}_i$ represents the DL channel from $ris_i$ to UE; $G_i$ represents the DL channel from NB to $ris_i$; $\mathbf{\Phi}_{ris,i}$ represents the regulation matrix of $ris_i$; $L$ represents the number of RIS participating in collaborative regulation.

Correspondingly, the received signal $\mathbf{Y}_{DL,CoMP-JT}$ at the UE side can be expressed as equation (6).

$$\mathbf{Y}_{DL,CoMP-JT} = \sum_{i=1}^{L} \mathbf{H}_i^H \mathbf{\Phi}_{ris,i} \mathbf{G}_i \mathbf{F} \mathbf{X} + \mathbf{N} \qquad (6)$$

where, $\mathbf{F}$ represents the precoding matrix at the NB side; $\mathbf{X} = \{x_1, x_2, x_3, ..., x_k\}^T$ represents a UE signal sequence with a length of $K$; $N$ represents the thermal noise matrix of the UE side.

Building upon the limited literature review, it becomes evident that while numerous articles have addressed the multi-RIS issue [16]-[18], there exists a gap in research concerning potential phase differences during joint transmission or reception in multi-RIS cooperation. In our previous CoMP research, we explored the phase difference issue arising from antenna reciprocity calibration between multiple Access Points (APs) in downlink CoMP-JT [20]. This phase difference poses challenges in achieving coherent CoMP-JT, thereby limiting the performance potential of the CoMP mechanism. In the context of multi-RIS collaboration, neglecting the careful consideration of phase relationships between different RISs may result in significant phase differences, rendering the realization of coherent CoMP-JT difficult. It is noteworthy that the absence of active Radio Frequency (RF) units in RIS eliminates the antenna reciprocity calibration problem discussed in [20]. The inherent challenge of antenna reciprocity calibration in traditional active phased array antennas stands as a primary constraint, making it challenging to realize coherent CoMP-JT and traditional CoMP in engineering. However, since RIS lacks the reciprocity calibration problem associated with active phased array antennas, achieving coherent CoMP-JT based on RIS becomes feasible. This possibility opens the door for realizing CoMP or Cell-free in future RIS networks.

However, in the scenario of multi-RIS collaboration, phase differences may still emerge due to the distinct path signal components of different RIS. The phase difference primarily comprises the following two factors (for illustrative purposes, we will consider a scenario involving two RIS).

The first factor stems from the fact that the electromagnetic wave components of the signal regulated by two RIS may traverse different propagation path lengths, leading to a phase difference denoted as $\Delta\varphi_1$.

The second factor lies in the fundamental functional characteristic of RIS, enabling the regulation of the propagation delay, phase, polarization, and frequency of the incident electromagnetic wave carrying the signal. When two RISs regulate the signal electromagnetic wave components of different transmission paths from UE, the signal electromagnetic wave components may exhibit varying phase differences with respect to the reference phase (typically based on the phase of the transmitter reference antenna). This discrepancy results in a phase difference denoted as $\Delta\varphi_2$ between the two signal electromagnetic wave components.

Then the total phase difference $\Delta\varphi_{total}$ is shown in Formula (7).

$$\Delta\varphi_{total} = \Delta\varphi_1 + \Delta\varphi_2 \qquad (7)$$

Since multiple cooperative RISs are usually located in a limited coverage area, it can be assumed that the path propagation delay does not exceed the CP length, and it will not lead to a large phase deviation. Therefore, the phase deviation of multiple RIS scenarios mainly comes from the phase difference between different RIS caused by RIS regulation. Therefore, Formula (7) can be approximately expressed as Formula (8).

$$\Delta\varphi_{total} \approx \Delta\varphi_2 \qquad (8)$$

Without losing generality, the downlink will be discussed as an example. According to the phase difference analysis of CoMP-JT in reference [14], assuming that the reference phase of the control matrix of multiple cooperative RIS are aligned with the phase of a same reference antenna $ant_{ref}$ of one NB, the regulation matrix of RIS can be expressed as formula (9) .

$$\boldsymbol{\Phi}_{ris,i} = c_{ris,i}\boldsymbol{\Phi}_{ris,i}^0 \qquad (9)$$

where, $c_{ris,i} = e^{-j\varphi_i}$ is a complex scalar, indicating the phase offset introduced by the $ris_i$ , relative to the reference antenna $ant_{ref}$ ; $\boldsymbol{\Phi}_{ris,i}^0$ represents the regulation matrix after $ris_i$ aligns the phase of the reference antenna $ant_{ref}$ ; Then the phase difference $\Delta c_{j,k}$ between two cooperative RISs, $ris_j$ and $ris_k$ , can be expressed as formula (10).

$$\Delta c_{j,k} = c_{ris,j}/c_{ris,k} = e^{-j(\varphi_j - \varphi_k)}$$
$$|\varphi_j - \varphi_k| \in [0, 2\pi) \qquad (10)$$

Therefore, after the regulation of two cooperative RIS, the phases of different path components of the signal arriving at UE are not aligned, which leads to the non-in-phase superposition of these multi-path signal components, thus reducing the received signal strength.

Let $p_{ue,i}$ denote the power allocated to $ue_i$ and $\sigma_{ue,i}$ denote the noise at the receiver of $ue_i$ . The signal to noise ratio (SINR) $\gamma_{ue,i}$ of $ue_i$ is then given by

$$\gamma_{ue,i} = \frac{\|\sum_{j=0}^{N-1}\mathbf{H}\boldsymbol{\Phi}_j + \mathbf{H}\|^2\, p_{ue,i}}{\sigma_{ue,i}} \qquad (11)$$

For NCM mode, NB can adjust phase alignment by coordinating multiple RISs, thus avoiding the phase difference between RISs. For example, the phase difference among RISs can be obtained by measuring the uplink signal components of a UE regulated by different RIS. Assuming that the uplink and downlink channels are reciprocal, NB can coordinate multiple RIS phase adjustments to perform phase alignment and collaboration regulation of RIS for uplink and downlink phase calibration. That is to say, in NCM mode, NB can skillfully control the regulation behavior of RIS, realize the phase alignment between RIS, and thus realize coherent joint transmission. In addition, the phase adjustment of multiple cooperative RIS can even be further coordinated to compensate the phase deviation caused by different propagation path lengths, thus reducing the influence of multipath.

Assuming two RISs, $ris_1$ and $ris_2$ , the independently calculated regulation matrices are $\boldsymbol{\Phi}_{ris,1} = c_{ris,1}\boldsymbol{\Phi}_{ris,1}^0$ and $\boldsymbol{\Phi}_{ris,2} = c_{ris,2}\boldsymbol{\Phi}_{ris,2}^0$. Network base station (NB) coordinates and updates the regulation matrices of $ris_1$ and $ris_2$ : keeping the regulation matrix $\boldsymbol{\Phi}_{ris,1}$ of $ris_1$ unchanged, and updating the $ris_2$ regulation matrix $\boldsymbol{\Phi}_{ris,2}$ to align with the phase of $ris_1$ . Based on equation (10), the updated regulation matrix $\widetilde{\boldsymbol{\Phi}}_{ris,2}$ of $ris_2$ can be expressed as equation (12).

$$\widetilde{\boldsymbol{\Phi}}_{ris,2} = (c_{ris,1}/c_{ris,2})\boldsymbol{\Phi}_{ris,2} \qquad (12)$$

For the SAM mode, due to the lack of network coordination constraints, each RIS usually determines the form of the regulation matrix to be used independently, which may lead to random phase differences between them. Referring to Equation (9), for a single RIS, the beam space direction of its regulation matrix $\boldsymbol{\Phi}_{ris,i}$ depends on the part $\boldsymbol{\Phi}_{ris,i}^0$ of the expression, and the coefficient $c_{ris,i} = e^{-j\varphi_i}$ can be any complex scalar with modulus of 1, which will not affect its beamforming performance. However, for multi-RIS collaborative regulation scenarios, the difference in complex scalar coefficients of different RIS reflects the phase relationship of beam vectors. Therefore, if the regulation matrix form of multiple collaborative RIS are not coordinated, the beams of each other may have random phase differences. The random phase difference of the regulation matrix between different RIS may lead to the beam phases misalignment of the two signal components arriving at the UE receiving antenna, thus leading to the non-phase superposition relationship of the signal components, which is called non-coherent CoMP-JT.

In the absence of NB unified coordinated control, there are two possible methods to solve the phase misalignment problem in SAM mode. The first method is to ignore the problem of phase deviation, and does not perform special calibration, which only supports incoherent CoMP-JT based on multiple RISs. However, this simple and rough method will greatly limit the performance of RIS networks. The second method is to constrain the phase regulation behavior of RIS itself, aligning it with the predefined reference phase $\varphi_0$. When calculating its regulation matrix, each RIS needs to constrain the form of the regulation matrix to make the complex scalar coefficient $c_{ris,i} = e^{-j\varphi_0}$. However, this method may bring two problems: on the one hand, it may limit the degree of freedom of RIS in electromagnetic wave adjustment, thus limiting its adjustment function in some scenes; On the other hand, there may be systematic deviation between RISs due to the deviations in the manufacturing process of RIS and system drift in the actual use environment. To minimize the system deviation caused by the latter, RIS needs to carry out necessary phase calibration before leaving the factory.

### B. MULTI-USER ACCESS

If multiple user equipments (UEs) are connected simultaneously in a community and their signals are incident on the same RIS panel, the RIS can regulate the signals of these UEs simultaneously. To match the regulation matrix of the RIS with the channels of each UE and achieve better performance, it is necessary to accurately measure and obtain the cascaded channel state information (CSI) of each UE passing through the RIS. Existing studies typically assume that the CSI of all UEs can be easily measured [15][16], and the regulation is optimized based on the CSI of multiple UEs. However, in practical deployment scenarios, measuring CSI to obtain multiple UE cascaded channels simultaneously usually faces many limitations.

Generally, CSI is obtained by estimating the reference signal/pilot signal of each UE. For multi-UE scenarios, it is necessary to identify and estimate the reference signals of multiple UEs respectively to obtain CSI of each UE. The ability to identify and estimate different UE reference signals/pilot signals is fundamental to obtaining CSI for multiple UEs.

For NCM mode, with the support of NB, the RIS can easily identify and measure the reference signals of multiple UEs, and obtain the CSI of each UE respectively. Therefore, the regulation matrix of the RIS can be collaboratively optimized to better match multiple UE channels, thereby achieving better performance.

For SAM mode, due to the lack of NB support, it is usually difficult for the RIS itself to distinguish signals from different UEs, and it is difficult to measure the CSI of each UE separately, making it difficult to achieve collaborative regulation for multiple UE access. Therefore, non-collaborative regulation can only be used in this case.

#### 1) Multi-UE Access with RIS Collaborative Regulation

Assuming that it is ideal to acquire CSI for each UE independently based on the NCM deployment mode, RIS adopts cooperative regulation to support multi-UE access. In this scenario, two types of algorithms will be explored to achieve the calculation of the RIS collaborative regulation matrix, including the pattern addition (PA) based multi-beam mechanism and RIS blocking mechanism.

#### (1) PA-based Multi-beam Supporting Multi-UE Access

According to the theory of multi-beam antenna, multi-beam can be formed by summing the required excitation coefficients of each beam at each antenna element [16]-[18]. Following this principle, we can generate multiple regulation beams in an RIS by adding the required regulation coefficients for different beam directions, which refers to PA technology [19]. The existing literature only explores the application of the PA mechanism to regulate multiple beams for simultaneous wireless energy transmission to multiple devices in RIS, and there is no published achievements to explore the application of the PA mechanism to multi-beam regulation in the RIS to support multi-user access [19][20]. Multi-device wireless energy transmission is significantly different from multi-user access. The former broadcasts wireless energy to multiple devices, while the latter transmits independent signals to multiple user devices. Therefore, it is necessary to further study the use of the PA mechanism for multi-beam regulation in the RIS to support multi-user access. In this case, we will discuss the support of PA mechanism for multi-user access. We will explore the specific methods and algorithms for generating multiple beams in a RIS using the PA mechanism for multiple UE. In addition, we will consider practical challenges, such as interference between users.

Based on the PA formula (14) in reference [19], the PA mechanism implements the RIS multi-beam regulation, which can be extended to support the simultaneous access of multiple users. The expression of the RIS regulation matrix $\mathbf{\Phi}_{mu}$ is formula (13).

$$\mathbf{\Phi}_{mu} = \sum_{k=1}^{K} \alpha_k \mathbf{\Phi}_k \tag{13}$$

where, $\mathbf{\Phi}_k$ represents the RIS regulation matrix corresponding to $ue_k$ independent access; the complex number $\alpha_k = \rho_k \exp(-j\vartheta_k)$ represents the weighting coefficient of the regulation matrix component of $ue_k$, which satisfies the constraint $\sum_{k=1}^{K} \|\alpha_k\|_2 \leq 1$; $K$ represents the total number of UE simultaneously accessed.

Taking $K$ users allocated to occupy the same time-frequency resources as an example,, the received signal $Y_{ue_k}$ of $ue_k$ can be expressed as formula (14).

$$\mathbf{Y}_{ue_k} = \mathbf{H}_k^H \mathbf{\Phi}_{mu} \mathbf{G}_k \mathbf{F}_k \mathbf{X}_{ue_k} + \sum_{i=1,i\neq k}^{K} \mathbf{H}_i^H \mathbf{\Phi}_{mu} \mathbf{G}_i \mathbf{F}_i \mathbf{X}_{ue_i} + \mathbf{N}_{ue_k} \tag{14}$$

where, $\mathbf{F}_k = f(\mathbf{H}_{ris,k}^H, \mathbf{\Phi}_{ris,k}, \mathbf{G}_k)$ represents the precoding matrix for $ue_k$ on the BS side; $\mathbf{X}_{ue_k} = [x_{ue_k,1}, x_{ue_k,2}, x_{ue_k,3}, ..., x_{ue_k,L}]^T$ represents the signal sequence of $ue_k$ , with a sequence length of $L$ ; $N_{ue_k}$ represents the Gaussian white noise matrix of $ue_k$ ; $\sum_{i=1,i\neq k}^{K}\mathbf{H}_i^H\mathbf{\Phi}_{mu}\mathbf{G}_i\mathbf{F}_i\mathbf{X}_{ue_i}$ represents the interference caused by other UE accessed simultaneously.

When using the PA mechanism to support multi-device wireless energy transmission, the received signal $Y_{dev_k}$ expression for device $dev_k$ is formula (15).

$$\mathbf{Y}_{dev_k} = \mathbf{H}_k^H\mathbf{\Phi}_{mu}\mathbf{G}_k\mathbf{F}_k\mathbf{S} \qquad (15)$$

where, $F_k$ represents the precoding matrix on the BS side for device $dev_k$ ; $\mathbf{S}=[s_1,s_2,s_3,...,s_L]^T$ represents a wireless signal sequence carrying wireless energy, with a sequence length of $L$ .

Comparing the two scenarios mentioned above (Formula 14 and Formula 15), it can be seen that the PA mechanism is used for broadcasting wireless energy from multiple devices without considering the Gaussian white noise matrix on the receiving side, and there is no interference from other devices. Whereas the PA mechanism supporting multi-UE access must consider the impact of the above two factors, special optimization of the algorithm is necessary.

From formula (13), it can be found that the RIS regulation matrix $\mathbf{\Phi}_{mu}$ using the PA mechanism is a weighted superposition of the control matrices obtained for each UE, which is the vector sum. The magnitude of the vector sum is related to the angle and the weighting coefficient between matrices. Therefore, by rotating the appropriate angle relationship and adjusting the appropriate weighting coefficient, it is natural to obtain that the PA mechanism can be optimized. Taking the channel capacity maximization criterion as an example, based on Formula (13), the RIS optimization regulation matrix $\mathbf{\Phi}_{mu,opt}$ is shown in Formula (16).

$$\mathbf{\Phi}_{mu,opt} = \underset{\vartheta_k\in(0,2\pi],0\leq\rho_k\leq 1}{agr\max} C(\sum_{k=1}^{K}\exp(-j\vartheta_k)\rho_k\mathbf{\Phi}_k) \qquad (16)$$

(2) RIS blocking mechanism Supporting Multi-UE Access

The RIS blocking mechanism operates by dividing the RIS antenna array into independent sub-groups, each with its own regulation matrix that can be configured separately [5]. These sub-groups are assigned to different UEs, and their corresponding regulation matrices are optimized and configured for each UE. This mechanism is used for multiple UE access when the UEs come from the same network, and for the coexistence of multiple networks when the UEs come from different networks. Although the RIS blocking mechanism can be used for multiple UE access, it also poses some challenges.

In multi-UE access scenarios using the RIS blocking mechanism, the signal beam broadening is often larger than the size of a single RIS sub-block, or even larger than the entire RIS panel. As a result, a UE signal beam may be incident on multiple sub-blocks simultaneously, and the signal components incident on different sub-blocks are regulated by their corresponding regulation matrices. While the signal components incident on its own RIS sub-block will be optimized and regulated, those incident on other sub-blocks may be unexpectedly and abnormally regulated.

Without loss of generality, consider a scenario where a RIS is divided into two sub-blocks, $ris_{sub1}$ and $ris_{sub2}$ , which are allocated to two UEs, $ue_1$ and $ue_2$ , respectively. The received signals $\mathbf{Y}_{ue_1}$ and $\mathbf{Y}_{ue_2}$ of $ue_1$ and $ue_2$ can be expressed as formulas (16a) and (17b):

$$\mathbf{Y}_{ue_1} = (\sqrt{\beta_{ue_1}}(\mathbf{H}_{ris\_sub1-ue_1}^H\mathbf{\Theta}_{ris\_sub1,ue_1}\mathbf{G}_{nb-ris\_sub1,ue_1}) $$
$$+ \sqrt{(1-\beta_{ue_1})}((\mathbf{H}_{ris\_sub2-ue_1}^H\mathbf{\Theta}_{ris\_sub2,ue_2}\mathbf{G}_{nb-ris\_sub2,ue_1}) \quad (17a)$$
$$+ \mathbf{H}_{nb-ue_1})\mathbf{F}_{ue_1}\mathbf{X}_{ue_1} + \mathbf{W}_{ue_1}$$

$$\mathbf{Y}_{ue_2} = (\sqrt{\beta_{ue_2}}(\mathbf{H}_{ris\_sub2-ue_2}^H\mathbf{\Theta}_{ris\_sub2,ue_2}\mathbf{G}_{nb-ris\_sub2,ue_2}) $$
$$+ \sqrt{(1-\beta_{ue_2})}((\mathbf{H}_{ris\_sub1-ue_2}^H\mathbf{\Theta}_{ris\_sub1,ue_1}\mathbf{G}_{nb-ris\_sub1,ue_2}) \quad (17b)$$
$$+ \mathbf{H}_{nb-ue_2})\mathbf{F}_{ue_2}\mathbf{X}_{ue_2} + \mathbf{W}_{ue_2}$$

where, $\beta_{ue_1} \leq 1$ and $\beta_{ue_2} \leq 1$ are the energy ratios of the $ue_1$ and $ue_2$ signals incident on their own RIS sub-block, $(1-\beta_{ue_1})$ and $(1-\beta_{ue_2})$ represent the energy ratios of $ue_1$ and $ue_2$ signals incident on RIS sub blocks which do not belong to these UE, $\mathbf{\Theta}_{ris\_sub1,ue_1}$ is the optimization regulation matrix of RIS sub-block $ris_{sub1}$ for the signal of $ue_1$ , and $\mathbf{\Theta}_{ris\_sub1,ue_2}$ is the optimization regulation matrix of RIS sub-block $ris_{sub2}$ A for the signal of $ue_2$ .

According to formula (16a), the signal of $ue_1$ has an $(1-\beta_{ue_1})$ proportion of energy falling on RIS sub-block $ris_{sub2}$ , where it is unexpectedly and abnormally regulated by $\mathbf{\Theta}_{ris\_sub1,ue_2}$ . $ue_2$ is like it and will not be repeated.

Therefore, it can naturally be concluded that an important factor in designing the RIS blocking mechanism is to ensure an appropriate number of RIS sub-blocks and the appropriate blocking ratio factor β . This needs to consider the trade-off between the signal broadening effect and the number of UE supported simultaneously.

*2) Multi-UE Access with RIS Non-Collaborative Regulation*
Assuming the SAM deployment mode is used and there is no NB support, it can be challenging for an RIS to distinguish signals from different UEs and measure the CSI of each UE separately. Therefore, non-collaborative regulation is the only way to enable multiple UE access. To address this issue, this section proposes two candidate solutions that involve adding a network entity with a simple structure and single function to perform multi-UE CSI measurements autonomously. Each solution is analyzed and discussed separately.

The first method is to get the regulation matrix $\mathbf{\Phi}_{mix}$ of RIS based on the superimposed mixed channels $\mathbf{H}_{mix}$ based on mixed signals from multiple UEs.

$$\mathbf{H}_{mix} = \sum \mathbf{H}_i \tag{18}$$

where, $H_i$ is the channel of $ue_i$.

This method is very simple without special optimization. However, it is obvious that the regulation matrix $\mathbf{\Phi}_{mix}$ will not match the channel of each UE well and will lead to performance degradation.

The second method is that RIS can only distinguish the signal of one UE (target UE), so it can only measure the channel $\mathbf{H}_{target}$ of the target UE, whereas the channel $\mathbf{H}_{non-target}$ of other UE (non-target UE) is unknown. Then, the RIS regulation matrix $\mathbf{\Phi}_{target}$ is calculated based on $\mathbf{H}_{target}$, which can well match the channel $\mathbf{H}_{target}$. Since different UE channels are independent, the regulation matrix $\mathbf{\Phi}_{target}$ is random beam for channels $\mathbf{H}_{non-target}$ of non-target UEs, which will lead to poor performance of non-target UE.

If the complexity of calculating the RIS regulation matrix is disregarded, NCM can assume that the CSI of all related UE can be ideally obtained, resulting in optimal performance. In practice, the complexity of implementation must be considered. One possible simple solution is to use the analog beam scanning method, similar to the hybrid beamforming used in massive MIMO, to scan the beam with a fixed period. The NB configures the RIS to perform beam scanning at a fixed interval, and the UE measures the RSRP of the corresponding beam and feeds it back to the NB. The NB then schedules the UE corresponding to the beam with high RSRP, meaning only the UE covered by the current beam is scheduled. From the perspective of a specific UE, only the time of corresponding beam scanning can be covered and scheduled. At other times, the scanning beam cannot be served because it cannot cover the UE. In other words, a UE can only be served through TDM mode, which greatly affects the UE's peak rate. Note that from a network throughput perspective, if there are a large number of UEs with good channel matching at each beam scanning time, then network throughput can be well guaranteed.

As an independent RIS cannot distinguish non-target signals, covering the non-target UE with a random beam or a beam of the target UE are possible ways to achieve this.

### C.  MULTI-CELL COORDINATION

In a traditional cellular network, the coverage area of a cell is limited by sector splitting to prevent the signal from exceeding a predetermined coverage area. This is done to limit signal strength at the edge of the cell and reduce interference levels to neighboring cells. In a traditional cell, only the NB is located at the center of the cell as a network unit. As long as the link budget from the NB to the cell edge is designed and the transmission power of the NB is allocated reasonably, the goal of limiting signal strength at the cell edge can be achieved effectively.

The introduction of RIS may bring about a new network topology, as shown in FIGURE 1. The RIS network topology includes not only the traditional network unit NB located at the cell center but also the new network unit RIS deployed in a distributed manner. In a typical scenario, RIS is deployed at the edge of the cell (FIGURE 1 a/b). RIS deployed at the cell edge can not only enhance the UE signal of the cell but also bring stronger interference to the UE at the edge of the neighbor cell [21][22]. For example, the DL beam regulated by RIS may point to neighbor cells, causing serious interference to the DL signal of UE in neighbor cells. For the UL beam regulated by RIS, the interference signal of neighbor cells can be shaped and aligned with NB, generating strong interference to the UL signal of the target cell. Taking the downlink signal as an example (FIGURE 2), the DL signal of the interfered neighbor cell can be expressed as Formula (18).

$$\mathbf{Y}_{DL,ue_1} = (\mathbf{H}_{DL,ris_1-ue_1}^H \mathbf{\Phi}_{ris_1,ue_1} \mathbf{G}_{DL,nb-ris_1} + \mathbf{H}_{DL,direct_1,ue_1}) \mathbf{F}_{ue_1} \mathbf{X}_{ue_1}$$

$$+ \sum_{i \neq 1} (\mathbf{H}_{DL,ris_2-ue_1}^H \mathbf{\Phi}_{ris_2,ue_i} \mathbf{G}_{DL,nb-ris_2} + \mathbf{H}_{DL,direct_2,ue_1}) \mathbf{F}_{ue_i} \mathbf{X}_{ue_i} + \mathbf{N} \tag{19}$$

where, $\mathbf{H}_{DL,ris_2-ue_1}$ is the interference channel between $ris_2$ of the neighbour cell to $ue_1$.

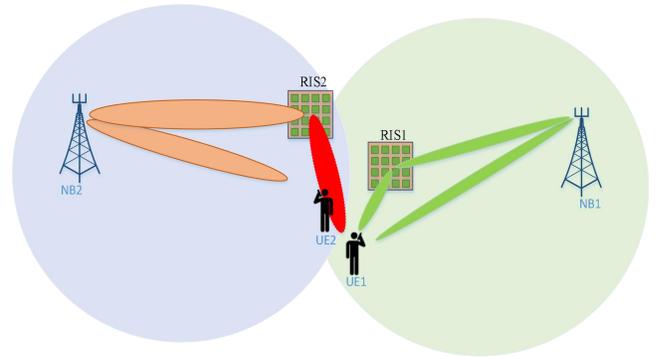

FIGURE 2. DL signal interfered by neighbour cells due to the introduction of RIS.

To solve the interference problem of neighboring cells caused by introducing RIS, there are several possible solutions as follows.

Scheme 1, the semi-static signal strength profile is predefined as covering the strength profile with cells from the center to the edge. This contour line is used to constrain the beamforming gain brought by the RIS regulation beam, thus restricting the interference to neighboring cells.

$$s\cos\vartheta \leq s_0 \qquad (20)$$

where, $\vartheta$ A signifies the angle between the beam normal direction of the RIS and the connection line linking the base stations of the target cell and its neighboring cells; $s_0$ represents the predetermined maximum threshold of signal strength.

Scheme 2, an extremely narrow beam is shaped, so the interference probability is low. In this case, the equivalent interference strength can be defined as: interference probability x signal strength.

$$I_e = p_I \times s \qquad (21)$$

$p_I$, interference probability;

$S$, signal strength;

$p_I = \dfrac{\theta}{2\pi} \leq p_{I,0}$, where $\theta \leq \theta_0$ represents the broadening angle of the beam, and the $\theta_0$ is representative of the predetermined broadening angle of the beam, as ascertained by the maximal threshold of interference probability $p_{I,0}$.

Scheme 3, By adjusting the beam forming gain, for example, to a wide beam, the interference to neighboring cells can be reduced. To predefine a set $\Phi$ of distinct beam widths contingent upon the placement of the RIS within the cell. Opt for beams of varying widths depending on the angle $\vartheta$ formed between the normal direction of the beam and the connecting line linking the base stations of the local and neighboring cells. In instances where the angle is minimal, opt for a broader beam, whereas in scenarios where the angle is substantial, opt for a narrower beam.

In NCM, the NB can intelligently manage the RIS regulation matrix calculation. For instance, when the NB detects potential interference at the edge of a neighboring cell close to the RIS, it can restrict the computation of the RIS regulation matrix to minimize interference to UEs in those neighboring cells. Conversely, if there are no UEs at the edge of adjacent cells, the RIS can optimize its control matrix without restrictions, maximizing its own performance.

In SAM, lacking NB support makes it challenging to obtain real-time information about interference from neighboring cells using only the RIS. One potential approach is to employ static methods to limit interference to neighboring cells, such as determining the RIS direction and beam adjustment during deployment planning. However, this static constraint mode constrains the flexibility of RIS deployment and phase control, hindering the realization of RIS gains.

## IV. NUMERICAL SIMULATION AND DISCUSSION

In the preceding sections, we introduced two modes of RIS network deployment and conducted a thorough analysis and discussion of their application in three typical collaboration scenarios. Building upon these analyses, this section employs Monte Carlo simulation methods to numerically simulate the aforementioned collaboration scenarios. To streamline the simulation process, we make the assumption that the direct path between the NB and UE is obstructed, allowing us to disregard the direct path signal. Consequently, the channel model corresponding to Equation (2) will be applied for the numerical simulations.

In each simulation, we assume that there are two networks, one NB for each network and one UE for each NB. The channel is assumed to be a far-field model. Unless specified otherwise, we assume that the RIS employs a UPA with $M = M_x \times M_y$; we fix $M_x = 20$ and increase $\mathbf{M}_y = [8,16,32,64]$; the NB adopts a UPA with $\mathbf{N} = \mathbf{N}_x \times \mathbf{N}_y$, $\mathbf{N}_x = 8$ and $\mathbf{N}_y = 4$; and the UE adopts $K = 1$. The carrier frequency is set to $f = 28GHz$. The normalized power is $p = 1$, and the variance of the Gaussian white noise is $\sigma = 3.16e-11$. Please note that the simulation results presented below are normalized to portray the relative relationship between different schemes, rather than providing absolute values.

### A. NUMERICAL SIMULATION OF RIS-BASED COMP-JT

As previously discussed, NCM, with the assistance of NB, can attain phase alignment among RISs, facilitating the coherent CoMP-JT involving multiple RISs. Conversely, in SAM, lacking NB support makes achieving phase alignment between multiple RISs challenging, leading to only non-coherent CoMP-JT. This section will create a scenario deploying two RISs in a cell, simulating both coherent CoMP-JT and non-coherent CoMP-JT supported by these two RISs to assess and contrast the performance of NCM and SAM.

This simulation assumes the deployment of two RISs within a single cell. To simplify the simulation design, it is assumed that both RISs experience the same large-scale channel fading but exhibit different small-scale channel fading. The phase difference of the two sub-channel components reaching the UE receiving antenna, after being regulated by the two collaborating RISs, is assumed to be uniformly distributed within the range of $\Delta\varphi \in (0, 2\pi]$. As a consequence, the phases of different path components of the signal reaching a UE are not aligned and do not superpose in the same phase, referred to as non-coherent JT. This results in a reduction in the received signal strength. FIGURE 3 illustrates the achievable data rate versus the number $M$ of RIS elements. The blue curve represents the ideal performance of coherent CoMP-JT, whereas the red curve shows the lower performance of non-coherent CoMP-JT. The comparison of the two simulation curves reveals that non-coherent CoMP-JT leads to a significant decline in performance.

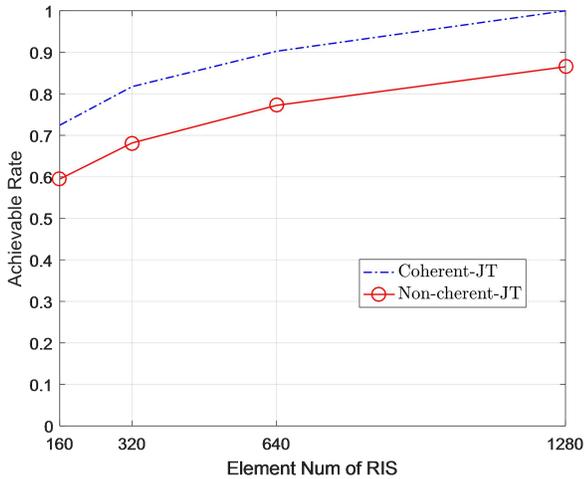

FIGURE 3. Performance of the Joint Transmission with Two RISs.

### B. NUMERICAL SIMULATION OF MULTI-UE ACCESS

From the analysis in section III, it is evident that the key difference between NCM and SAM is whether there is NB support to accurately obtain the relevant channel state information. To simplify the simulation, it is assumed that CSI of multiple UEs is perfect for NCM. In contrast, SAM has three cases: case 1, the mixed channel of multiple UEs can be obtained; case 2, the channel of UE is difficult to obtain; and case 3, only the channel of the first UE can be obtained, while the channels of other UEs are unknown. In the simulation, two UEs with Orthogonal Frequency-Division Multiple Access (OFDMA) are assumed, and their signals are regulated by the same RIS. The detailed simulation configuration information and results are presented below.

#### 1) Multi-UE Access Performance Using RIS Collaborative Regulation: PA Mechanism

This simulation assumes a far-field channel model, with the RIS configured as a uniform linear array (ULA), $\mathbf{N} = \{64, 128, 256, 512, 1024\}$, and the NB configured as a uniform linear array (ULA), $M = 64$,. The number of UEs is $K = 2$, and the carrier frequency is set to $f = 5GHz$. Using formulas (12) and (15) from section III, the simulation evaluates the multi-UE access performance and optimized PA performance of conventional PA mechanisms, providing performance comparisons for ideal regulation, unexpected regulation, and random phase regulation, respectively.

The performance curve in FIGURE 4 shows that the PA mechanism can provide better RIS collaborative regulation of multi-UE access performance. Furthermore, the performance of the PA mechanism can be further optimized through the phase relationship and weighting coefficients between the superimposed regulation matrices.

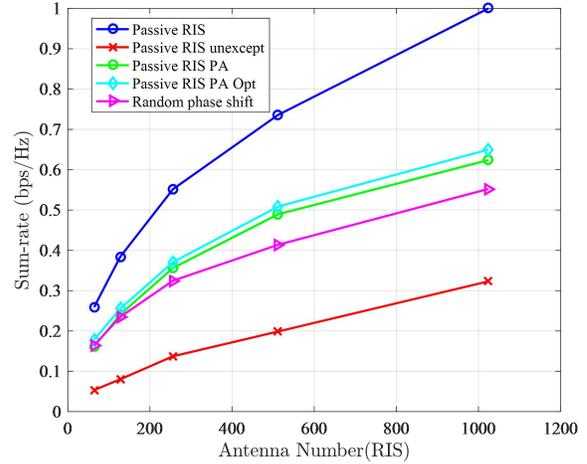

FIGURE 4. Multiple UE Access Performance Based on PA Mechanism RIS Collaborative Regulation ("Passive RIS" represents the most ideal performance without considering RIS regulation constraints; "Passive RIS unexpected" represents the performance that is unexpected to be regulated by another UE's regulation matrix; "Passive RIS PA" represents the performance when using PA mechanism but not optimized; "Passive RIS PA Opt" represents the performance after using PA mechanism and optimizing phase and weighting coefficients; "Random phase shift" represents the performance when using random beam regulation.)

#### 2) Multi-UE Access Performance Using RIS Collaborative Regulation: Blocking Mechanism

According to formula (16) of the RIS blocking mechanism, the regulation coefficient matrix of the first sub-block of the RIS matches the channel of the target signal ($ue_1$), whereas the regulation coefficient matrix of the second sub-block matches the channel of another target signal ($ue_2$). The target signal ($ue_1$) is optimized and regulated when incident on the first RIS sub-block, whereas it is unexpectedly and abnormally regulated when incident on the second sub-block.

In the simulation execution, different blocking proportion coefficients were simulated, namely the energy ratio incident on multiple sub-blocks. Without loss of generality, we can assume that the energy ratio $\beta$ of the signal incident on the two sub-blocks are equal to the proportion of the block size. The proportion coefficient of the signal energy of the unexpected regulation is $\beta$, and the value of $\beta$ is between 0 and 1, $\beta \in \{0.1, 0.2, 0.4, 0.5, 0.6, 0.8, 0.9\}$. FIGURE 5 and FIGURE 6 depict the relationship between the achievable data rate and the number $M$ of RIS units, with fixed transmission power $p$ and variable $M$, respectively. Assuming that the RIS performs regulation on non-target signals throughout all time slots.

From the curve in FIGURE 5, the proposed mechanism can greatly improve the performance of non-target signals, but it will also reduce the performance of target signals to a certain extent. The reason is that the RIS blocking mechanism reduces the effective antenna aperture of the target signal, and the RIS sub-blocks allocated to other users will also regulate the energy components of the target signal with unexpected regulation. However, as shown in FIGURE 6, the sum data rates of target UE and non-target UE (i.e., target signal and non-target signal) indicate that this RIS blocking mechanism can achieve higher sum rates. Therefore, it is necessary to balance the performance between target and non-target signals by selecting appropriate partition coefficients.

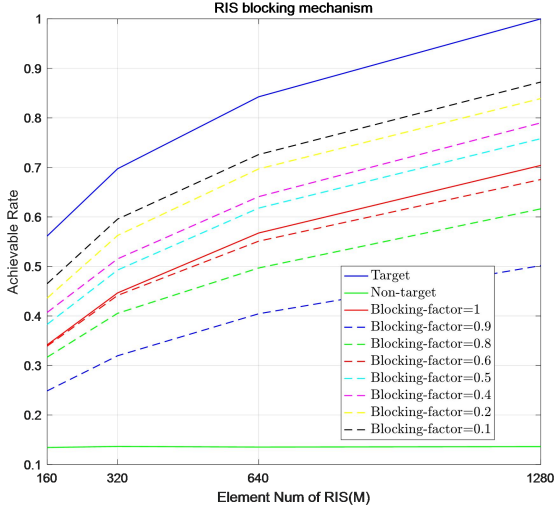

**FIGURE 5.** Multiple UE Access Performance Using Blocking Mechanism with collaborative regulation. . ("Target": RIS uses perfect CSI regulation signal; "Non-target": RIS uses traditional conventional RIS to unexpectedly regulate incident electromagnetic waves; "Blocking factor = {0...., 0.1}": RIS block mechanism is used with the block ratio coefficient)

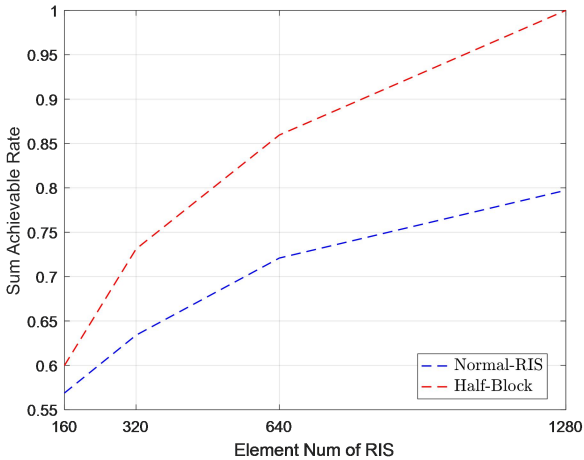

**FIGURE 6.** Reachability Rate of Target UE and Non-Target UE. ("Normal RIS": Assuming that two UEs use the full RIS panel; "Half block": blocking ratio 1:1)

### 3) Multi-UE Access Performance Using RIS Non-Collaborative Regulation

Based on the analysis in section above, this section will conduct numerical simulation evaluation and comparison of multi UE access performance for non-cooperative regulation of RIS in the following four scenarios.

Case 1, the CSI of one UE ( $ue_1$ ) is perfectly obtained;

Case 2, the regulation matrix $\Phi_{ue,2}$ of $ue_2$ is calculated based on the mixed channel of $ue_1$ and $ue_2$ ;

Case 3, the regulation matrix $\Phi_{ue,1}$ of RIS is calculated based on the CSI of $ue_1$ and used for $ue_2$ ;

Case 4, the random phase matrix is used for $ue_2$ .

FIGURE 7 depicts the achievable data rate versus the number $M$ of RIS elements. The curves with the legend of "UE1: Perfect CSI", "UE2: Random phase", "UE2: Mixed channel" and "UE2: Using UE1 CSI" correspond to the performance of the above case 1~4, respectively. These simulation results show that the performance of SAM is significantly reduced due to the inability to obtain CSI of multiple UE. Note that the "achievable rate" in Figure 7 represents the achievable rate of a single UE.

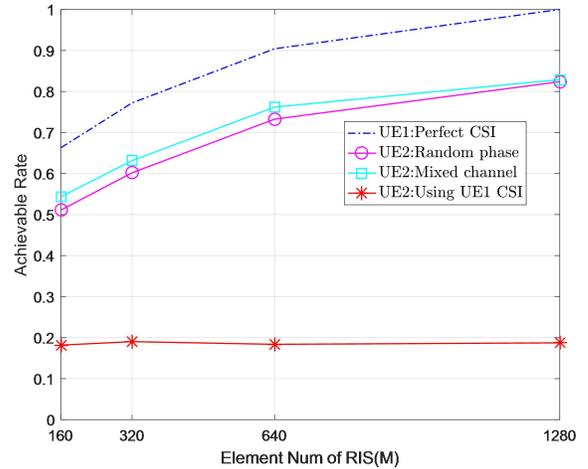

**FIGURE 7.** Performance of one RIS serving two UEs at the same time.

### C. NUMERICAL SIMULATION OF MULTI-CELL COORDINATION

The simulation assumes that there is a UE at the cell edge, and its signal is enhanced by the $ris_1$ of the serving cell, but it is also interfered by the interference signal of the neighboring cell enhanced by the $ris_2$ . FIGURE 8 depicts the achievable data rate versus the number $M$ of RIS elements. The curves in FIGURE 8 show three cases.

Case 1 (refer to the curve with the legend of "No-interference"), the interference signal of the neighboring cell is regulated to the null space of the interference channel by RIS, so the interference to the neighboring cell UE is 0 (corresponding to the upper bound of the performance);

Case 2 (refer to the curve with the legend of "Interference w/ beam"), the interference signal of the neighboring cell is perfectly matched with the interference channel after RIS regulation so that the interference of the UE to the neighboring cell reaches the maximum (corresponding to the lower bound of the performance);

Case 3 (refer to the curve with the legend of "Random-phase"), the interference of neighboring cells is regulated with random phase by RIS, resulting in a random interference effect.

In the first case mentioned above, the perfect CSI of the interference channel is obtained with the help of the NB in the NCM mode, and the interference signal of the neighboring cell is regulated to the zero space of the interference channel based on the CSI. The second and third cases refer to that in SAM mode, the CSI of the interference channel cannot be obtained without the assistance of NB, so it is impossible to coordinate and suppress the interference to the neighboring cells, resulting in serious interference. Because UE channels of two cells are independent, the third kind of random interference effect will be generated in general. The second situation reflects the worst case, that is, the lower bound of the performance of interference to neighbour cells after introducing RIS.

Comparing the performance of the UE with or without RIS-enhanced interference from neighboring cells, we can see that the performance of the UE has deteriorated seriously.

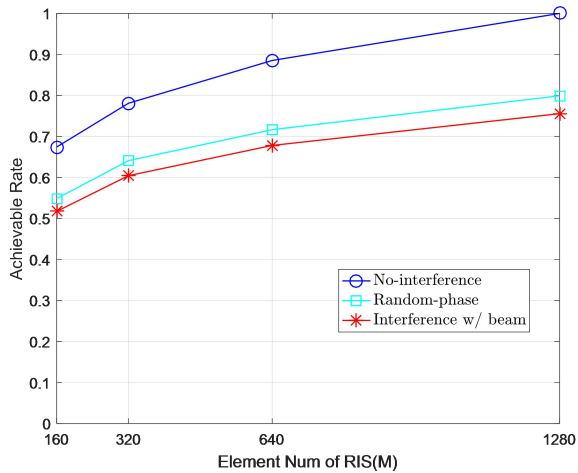

FIGURE 8. The performance of a UE with or without RIS enhanced interference from neighboring cells.

## V. FIELD TEST AND ANALYSIS

Due to constraints in implementation conditions, only a subset of the aforementioned scenarios has been field-tested, with simplified test conditions. Consequently, this paper refrains from directly comparing these field test results with the earlier simulation outcomes. Instead, the focus lies on the comparative analysis of the test results obtained under various configurations within the field test itself. This section presents field tests conducted in two typical indoor scenarios: coverage performance tests in an indoor L-shaped corridor and an open office area utilizing mmWave. For additional field test results in various RIS scenarios, readers are directed to the previously published article [23].

The field test performance metrics encompass the Reference Signal Received Power (RSRP) and downlink data rate. RSRP, a widely adopted standard in 5G systems, serves to assess reception quality at the receiver, as stipulated in [24]. RSRP is defined as the linear average over the power contributions of the resource elements carrying reference signals, with the synchronization signal block (SSB) employed as the reference signal in the subsequent simulations and tests. The general unit for RSRP is dBm per resource element (dBm/RE), and in the absence of an expected signal, the thermal noise measured at the UE is approximately -115 dBm/RE.

The RIS prototype, 5G-NR Node B (NB), and terminal were developed by ZTE Corporation. The 5G-NR NB and terminal are designed in accordance with 5G standards, and their details will not be expounded upon here. Focusing on the RIS (depicted in FIGURE 9), its operational bandwidth spans 1 GHz centered around a frequency of 27 GHz. The RIS adopts a Uniform Planar Array (UPA) structure featuring a $128 \times 128$ element grid, where the reflection coefficient can be adjusted using PIN diodes. These PIN diodes enable the configuration of phase shifts for the reflecting elements at $0\circ$ or $180\circ$. Additionally, each element incorporates two PIN diodes corresponding to horizontal and vertical polarization, resulting in four possible states for each element. Refer to Table 1 for more detailed parameters of the RIS.

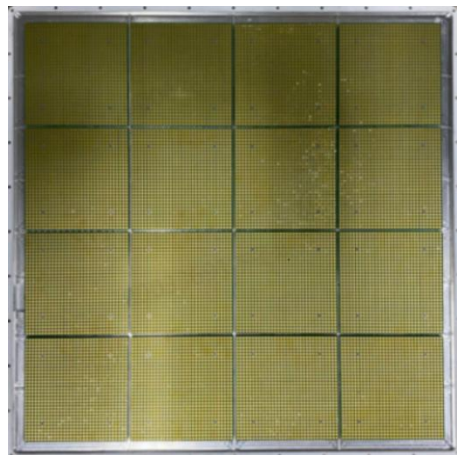

FIGURE 9. The RIS fabricated for the field test

TABLE I
PARAMETERS OF ZTE RIS USED IN FIELD TEST

| Parameters | Values |
|---|---|
| RIS Type | Reflective RIS |
| Center frequency | 27GHz |
| Effective working frequency range | 26-28GHz |
| Polarization mode | Bipolarization |
| Number of RIS antenna arrays | 16384 |
| Row number N and column number M of RIS antenna array | N=128, M=128 |
| The row number N and column number M of the regulatable RIS antenna array | N=128, M=128 |
| Phase range of RIS beam scanning (degrees) | 60/75 |
| RIS beam width(degrees) | 2~6 |
| Quantized bit number | 1 |
| Antenna array aperture of RIS | 660mm |

## A. FIELD TEST OF COVERAGE PERFORMANCE IN INDOOR L-SHAPED CORRIDOR SCENARIOS

Non-Line-of-Sight (NLOS) coverage within an indoor "L"-shaped corridor stands out as a representative indoor scenario, particularly in the mmWave frequency band. Consequently, the L-shaped corridor scenario has been chosen to establish an NLOS environment for assessing the enhancement of network performance following RIS deployment. The test objectives in this scenario encompass evaluating the coverage performance improvement facilitated by RIS deployment, contrasting the impact of RIS deployment with that of metal plates, and comparing the network coverage performance of RIS in NCM versus SAM. Detailed parameters of the 5G-NR Node B and terminal for the field test in the indoor L-shaped corridor are provided in Table 2.

TABLE II
CONFIGURATION PARAMETERS OF 5G-NR NODE B AND TERMINAL IN THE FIELD TEST OF INDOOR L-SHAPED CORRIDOR

| Parameters | Values |
|---|---|
| Center frequency | 27GHz |
| Bandwidth of base station | 400MHz |
| Transmission power of base station | 23dBm |
| Antenna gain of base station | 28dBi |
| Distance from base station antenna to RIS center point | 20m |
| Antenna gain of terminal | 9dBi |
| Transmitting power of terminal | 23dBm |
| Number of transceiver antennas of the base station | 4T4R |
| Number of transceiver antennas of the terminal | 2T2R |

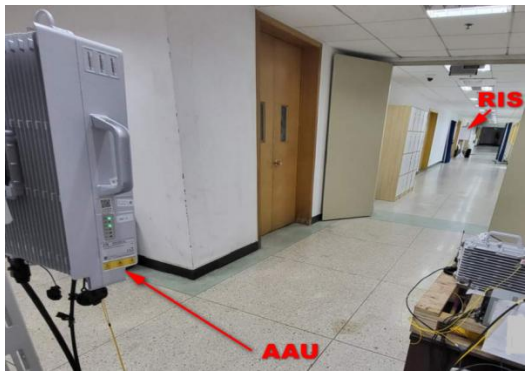

(a) The considered field test setup of the RIS-aided wireless communication system in an L-shaping corridor of indoor office environment.

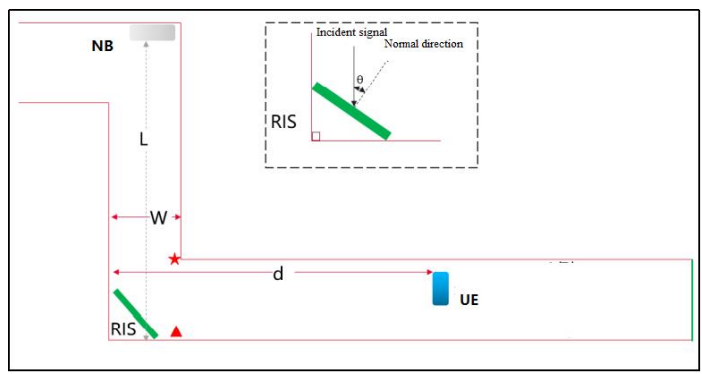

(b) the top view of the setup with all terminal test locations.

FIGURE 10. Field Test of RIS Coverage Performance in Indoor L-shaped Corridor Scenarios.

The considered system architecture of our RIS-assisted test setup is illustrated in FIGURE 10(a). The test activities are carried out in an L-shaped corridor in the indoor office environment, with the help of a 5G-NR Node B (NB) for transmission and a 5G-NR terminal (UE) for receiving air signals. In the test setup, the RIS is mounted on the hallway's corner, the NB is fixed and installed at the position shown in FIGURE 10(b) (at one side of the L-shaped corridor, different from the moving path of UE), whereas the UE is positioned at various distances from the center of RIS throughout the test process. The nearest point (test point #1) between the terminal and the RIS board is 2 meters, and every 2 meters along one side of the L-shaped corridor is used as a test point. In other words, the test points #1 to #6 are arranged in order from the nearest to the farthest. The angle of signal incidence to the RIS panel is 45 degrees. In addition, the cases of the power-off RIS and the metal plate with the same size as RIS are added as the comparison test items, and the case without RIS is taken as the test baseline.

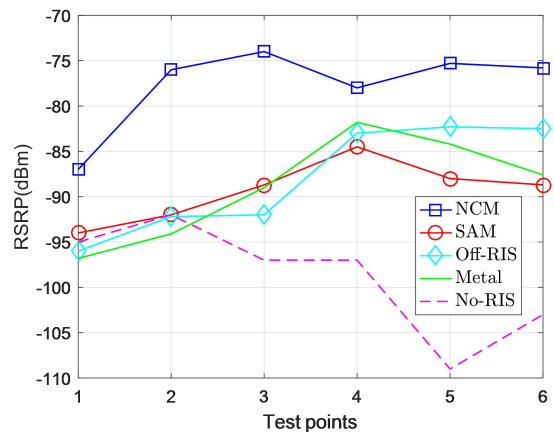

(a) RSRP

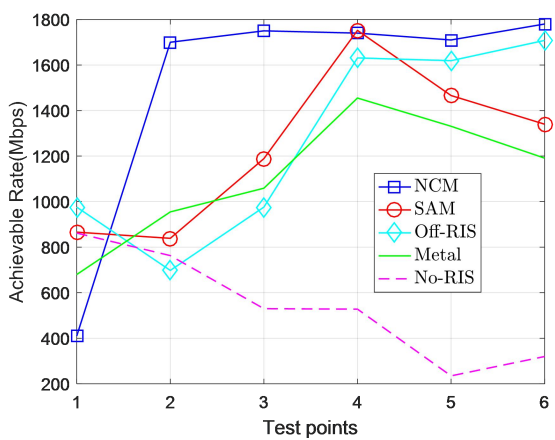

(b) Achievable Rate of Downlink

**FIGURE 11.** Field test results of the downlink in the Indoor L-shaped corridor scenarios ("NCM" refers to the performance curve of NCM, "SAM" is the performance curve of SAM, "Off-RIS" refers to the performance curve when the RIS is powered down, "Metal" refers to the performance curve when the metal plate is used, "No-RIS" refers to the performance curve when RIS is not deployed.)

During the test, the SAM utilizes a fixed RIS regulation matrix (codebook). The codebook is chosen to be appropriate for the channels encountered by the mobile terminal at various test points. In NCM, the RIS dynamically tracks the channel variations of the UE and selectively picks the RIS regulation codebook that optimally aligns with the UE channel. FIGURE 11 illustrates the RSRP and DL achievable rate variations of the terminal across these test points in both scenarios.

### B. FIELD TEST OF RIS COVERAGE PERFORMANCE IN OPEN AREA SCENARIOS

An open office area represents another common scenario for indoor wireless network coverage. The test objectives in this scenario primarily involve contrasting the network performance of two RIS deployment modes, NCM and SAM. Simultaneously, a metal plate of identical size to the RIS is introduced as a comparative test element, and the scenario without the presence of an RIS is considered the baseline for testing. Refer to Table 3 for detailed parameters of the 5G-NR Node B and terminal in the field test of the indoor open office area.

TABLE III
Configuration parameters of 5G-NR Node B and terminal in the field test of indoor open office area

| Parameters | Values |
|---|---|
| Center frequency | 27GHz |
| Bandwidth of base station | 400MHz |
| Transmission power of base station | O2I: 23dBm I2I: 19dBm |
| Antenna gain of base station | 28dBi |
| Distance from base station antenna to RIS center point | 20m |
| Antenna gain of terminal | 9dBi |
| Transmitting power of terminal | 23dBm |
| Number of transceiver antennas of the base station | 4T4R |
| Number of transceiver antennas of the terminal | 2T2R |
| Height of base station antenna | 1.8m |
| Office area size (L x W x H) | 9m*6.6m*4m |
| Interval size of test points | 1m |
| Distance from indoor base station to RIS # 1 | 2.1m |
| Distance from indoor base station to RIS # 2 | 4.2m |
| Distance from indoor base station to RIS # 3 | 3.8m |
| Distance from outdoor base station to RIS # 1 | 12.8m |
| Distance from outdoor base station to RIS # 2 | 16.2m |
| Distance from outdoor base station to RIS # 3 | 15.6m |

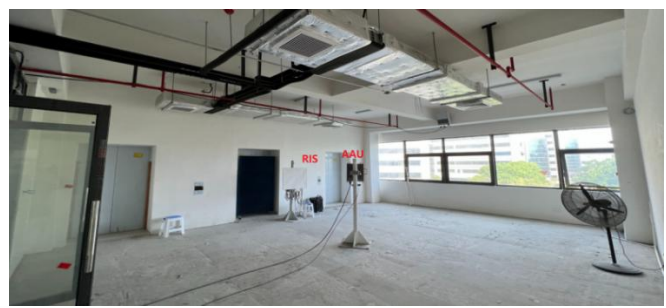

(a) The considered field test setup of the RIS-aided wireless communication system in an indoor open office environment

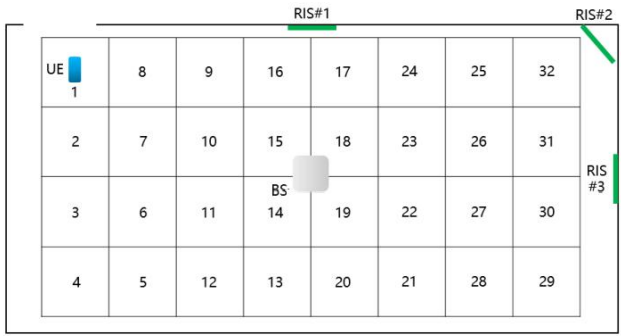

(b) the top view of the setup with all terminal and RIS test locations.

 **FIGURE 12.** Field Test of RIS Coverage Performance in Open Office Area Scenarios.

The system architecture of the RIS-assisted test setup is depicted in FIGURE 12. The experimental activities are conducted within an indoor open office environment utilizing a 5G-NR Node B for transmission and a 5G-NR terminal (UE) for receiving air signals. In the test configuration shown in FIGURE 12, the RIS is strategically placed at multiple locations, the Node B remains fixed in the center of the open office area, while the UE is positioned at various randomly selected points marked with numbers throughout the testing procedure. For the scenario involving a fixed beam (i.e., SAM case), the RIS beam is configured to target point 4 in FIGURE 12.

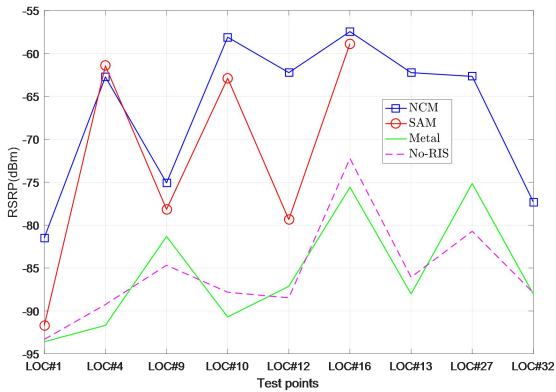

(a) RSRP

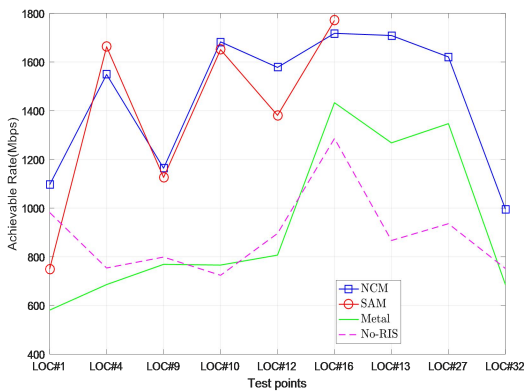

(b) Achievable Rate of Downlink

**FIGURE 13.** Field test results of the downlink in the Indoor open office scenarios ("NCM" is the performance curve of NCM, "SAM" is the performance curve of SAM, "Metal" refers to the performance curve when the metal plate is used, "No-RIS" refers to the performance curve when RIS is not deployed.)

In the experiment, a predetermined RIS regulation matrix (codebook) is employed in the SAM. The codebook is selected to direct the beamforming toward point #4 in FIGURE 12. In NCM, the RIS dynamically tracks the channel variations of the UE and flexibly chooses the RIS regulation codebook that optimally aligns with the UE channel. FIGURE 13 illustrates the fluctuations in RSRP and DL achievable rate of the terminal across the designated test points for these scenarios, respectively.

## C. FIELD TEST RESULTS ANALYSIS

The results depicted in FIGURE 11 and FIGURE 13 illustrate that in SAM, the RSRP and achievable rate change in response to the UE moving to different locations. When the channel aligned with the UE's location matches the regulation codebook of the RIS, the performance is relatively favorable. Conversely, in NCM, the RIS can consistently adaptively track channel changes and select the best-matching regulation codebook, resulting in consistently high RSRP and achievable rate, irrespective of the UE's location. Observations reveal that the RSRP and achievable rate in SAM are consistently lower than in NCM, with some points in close proximity. This discrepancy arises from SAM mode, where the adjustment codebook of the (RIS is not dynamically based on the best match of the current CSI. Instead, it achieves better performance by moving to the position where the channel closely aligns with the pre-determined codebook, lacking the adaptive channel matching seen in NCM. Additionally, in indoor scenarios, deploying metal panels or unpowered RIS panels can provide some gain compared to scenarios without RIS panels. However, due to the absence of programmable adjustment capabilities for metal panels and unpowered RIS panels, certain gains are limited to specific angles.

## VI. FUTURE TRENDS AND CHALLENGES

To further advance the industrialization process of RIS, certain aspects require additional in-depth research and standardization.

## A. COMPLEXITY AND OPTIMIZATION OF RIS NETWORKS DEPLOYMENT

The introduction of Reconfigurable Intelligent Surfaces (RIS) heralds a new paradigm for future networks, albeit accompanied by challenges in network deployment and optimization.

1) Multiple Transmission Nodes Sharing the Same RIS:
   - The coexistence of multiple transmission nodes sharing a common RIS introduces complexities in coordination and resource allocation.
1) Control Link Between Transmission Node and RIS:

- Establishing and managing control links between transmission nodes and RIS pose challenges, necessitating effective communication and synchronization.

2) Challenges in Network Topology Planning and Optimization:

- RIS's capacity to enhance and expand signal coverage challenges traditional sector-based coverage characteristics, introducing complexity to network planning and optimization efforts.

3) Site Selection and Power Supply:

- The simplicity, cost-effectiveness, and low power consumption of RIS enable widespread deployment. However, this ubiquity poses challenges in power supply and site selection for RIS installations.

4) Diverse Scenarios Requiring Varied Forms of RIS:

- Different scenarios demand different configurations of RIS. Therefore, the design and development of diverse forms of RIS become imperative to address the specific needs of various deployment scenarios.

*B. STANDARDIZATION AND PROTOCOL*

In this section, we delve into the standardization of RIS within the context of wireless cellular networks, excluding considerations of RIS standardization in other systems. Acknowledged as a prospective key technology for future wireless networks, numerous regional standardization organizations have initiated pertinent research and promotional activities for RIS. Anticipated integration into the 5G-Advanced and 6G standards is underway [25].

Evidently, NCM is poised to become the predominant deployment mode for RIS in 6G networks. To facilitate the application of NCM, the following three aspects of standardization should be considered:

1) Feedback of CSI and Scheduling of Beamforming:

- Addressing challenges associated with the substantial pilot costs and channel information feedback costs arising from the deployment of super-large antenna arrays.
- Meeting real-time requirements for channel measurement and feedback.

2) Standardizing the Electrical Interface of the Network Control Link:

- Establishing standards for the electrical interface of the network control link between the network and RIS. The deployment of this control link necessitates special considerations, introducing complexity and cost implications, particularly in remote areas.

3) Standardizing the Interaction Process of Control and Measurement Signaling:

- Standardizing the protocol governing the interaction process of control and measurement signaling. This standardization is essential for ensuring a coherent and efficient exchange of information between the network and RIS.

Analysis from preceding chapters indicates that the deployment of pure autonomous control-based SAM is unsuitable for 6G networks. However, in specific straightforward application scenarios, a variant of SAM with minimal network control can be employed to mitigate the need for extensive RIS network control. For instance, leveraging statistical CSI can reduce feedback overhead, and adjusting control granularity can alleviate complexity, thereby easing the deployment challenges associated with RIS networks.

## VII. CONCLUSION

This article conducts a comprehensive analysis and comparison of the two deployment modes in RIS-assisted wireless communications, namely NCM and SAM. It introduces three typical collaboration scenarios in RIS networks — multi-RIS collaboration, multi-user access, and multi-cell coordination—to highlight distinctions between the two deployment modes. The application of these modes in the given scenarios is thoroughly analyzed and compared, supported by established simulation models featuring rich numerical simulation results. Furthermore, a field test environment is established to conduct preliminary tests on both modes.

Theoretical analysis, numerical simulation, and field tests collectively demonstrate that while SAM offers the advantage of simple implementation, its performance lags behind NCM in the three typical collaboration scenarios. However, SAM's performance remains robust in simple single-user scenarios, reflecting its characteristics of simplicity and ease of implementation.

Consequently, two key observations emerge. Firstly, NCM proves suitable for complex networks and licensed spectrum environments with high coexistence requirements, such as cellular networks. Secondly, SAM is apt for simple networks and unlicensed spectrum technologies with local coverage, such as Wi-Fi. Realizing the engineering application of NCM and SAM necessitates the exploration of technical challenges and active promotion of standardization in the future.

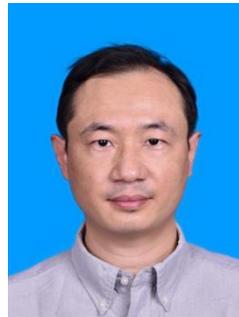

**Yajun Zhao** holds Bachelor's, Master's, and Doctoral degrees. Since 2010, he has assumed the role of Chief Engineer within the Wireless and Computing Product R&D Institute at ZTE Corp. Prior to this, he contributed to wireless technology research within the Wireless Research Department at Huawei. Currently, his primary focus centers on 5G standardization technology and the advancement of future mobile communication technology, particularly 6G. His research pursuits encompass a broad spectrum, including reconfigurable intelligent surfaces (RIS), spectrum sharing, flexible duplex, CoMP, and interference mitigation. He has played an instrumental role in founding the RIS TECH Alliance (RISTA) and currently holds the position of Deputy Secretary General within the organization. Additionally, he is a founding member of the RIS task group under the purview of the China IMT-2030 (6G) Promotion Group, where he serves as the Deputy Leader.